\documentclass[sigconf]{acmart}

\usepackage{balance}

\AtBeginDocument{%
  \providecommand\BibTeX{{%
    \normalfont B\kern-0.5em{\scshape i\kern-0.25em b}\kern-0.8em\TeX}}}

\setcopyright{acmcopyright}
\copyrightyear{2020}
\acmYear{2020}
\acmDOI{10.1145/1122445.1122456}

\acmConference[NFW '20]{New Future of Work '20: Symposium on New Future of Work}{August 03--05, 2020}{Online}
\acmBooktitle{NFW '20: Symposium on New Future of Work,
  August 03--05, 2020, Online}
\acmPrice{15.00}
\acmISBN{978-1-4503-XXXX-X/18/06}

\begin{document}

\title{Seating preference analysis for hybrid workplaces}

\author{Mohammad Saiedur Rahaman}
\affiliation{%
  \institution{RMIT University}
  \streetaddress{GPO Box 2476}
  \city{Melbourne}
  \state{VIC}
  \postcode{3000}
  \country{Australia}
}\email{saiedur.rahaman@rmit.edu.au}

\author{Shaw Kudo}
\affiliation{%
  \institution{Arup}
  \city{Melbourne}
  \state{VIC}
  \postcode{3000}
  \country{Australia}
}\email{shaw.kudo@arup.com}

\author{Tim Rawling}
\affiliation{%
  \institution{Arup}
  \city{Melbourne}
  \state{VIC}
  \postcode{3000}
  \country{Australia}
}\email{tim.rawling@arup.com}

\author{Yongli Ren}
\affiliation{%
  \institution{RMIT University}
  \streetaddress{GPO Box 2476}
  \city{Melbourne}
  \state{VIC}
  \postcode{3000}
  \country{Australia}
}\email{yongli.ren@rmit.edu.au}

\author{Flora D. Salim}
\affiliation{%
  \institution{RMIT University}
  \streetaddress{GPO Box 2476}
  \city{Melbourne}
  \state{VIC}
  \postcode{3000}
  \country{Australia}
}\email{flora.salim@rmit.edu.au}

\renewcommand{\shortauthors}{Rahaman, et al.}

\begin{abstract}

Due to the increasing nature of flexible work and the recent requirements from COVID-19 restrictions, workplaces are becoming more hybrid (i.e. allowing workers to work between traditional office spaces and elsewhere including from home). Since workplaces are different in design, layout and available facilities, many workers find it difficult to adjust accordingly. Eventually, this impacts negatively towards work productivity and other related parameters including concentration, stress, and mood while at work. One of the key factors that causes this negative work experience is directly linked to the available seating arrangements. In this paper, we conduct an analysis to understand various seating preferences of 37 workers with varying demographics, using the data collected pre-COVID-19, and analyse the findings in the context of hybrid workplace settings. We also discuss a list of implications illustrating how our findings can be adapted across wider hybrid work settings.

 
\end{abstract}


\maketitle

\section{Introduction and Background}

The concept of hybrid workplace is to remove physical barriers among teams by allowing employees to choose to work however and wherever they feel most productive. Workers in a hybrid setting are a mix between co-located and remote workers, (i.e., they spend part of their time working in the traditional office and part of their time working elsewhere) \cite{Colin1}. This trend shift in work arrangements is typically associated with a drive towards business efficiency: saving space, reducing costs and accommodating growing teams \cite{9097829}. At the same time, prior research highlights additional employee-focused benefits associated with hybrid work settings when workers share their office spaces. This includes greater worker satisfaction \cite{doi:10.1080/17508975.2012.695950,eric:academy,KIM201318} and more flexibility in the way they perform their work \cite{doi:10.2307/3556639,doi:10.1177/0018726709342932}. The shared workplace also has the potential to enable the exchange of knowledge more effectively among colleagues~\cite{Meijer:TF,doi:10.1002/job.1973,KIM201318}, which can enrich their respective skills and potentially lead to greater work productivity \cite{doi:10.1080/17508975.2012.695950}. However, it must be acknowledged that several studies have identified a number of problems \cite{10.1371/journal.pone.0193878,MORRISON2017103} associated with shared offices such as increased distrust, distractions, and uncooperative behavior. To measure the success of hybrid workplaces, both active and passive sensing have been used to monitor employees' organizational behavior \cite{4694078} and workplace interactions \cite{Brown:2014:TSI:2531602.2531641,GHAHRAMANI201842,Mark:2014:CMF:2531602.2531673,Brown:2014:AIT:2632048.2632056, Mashhadi:2016:CSC:2957265.2957272}. The outcomes of this research can help better understand the emotions, productivity and team dynamics of employees, which could lead towards overall organizational success.

\begin{figure*}[h]
  \centering
  \includegraphics[width=0.9\textwidth]{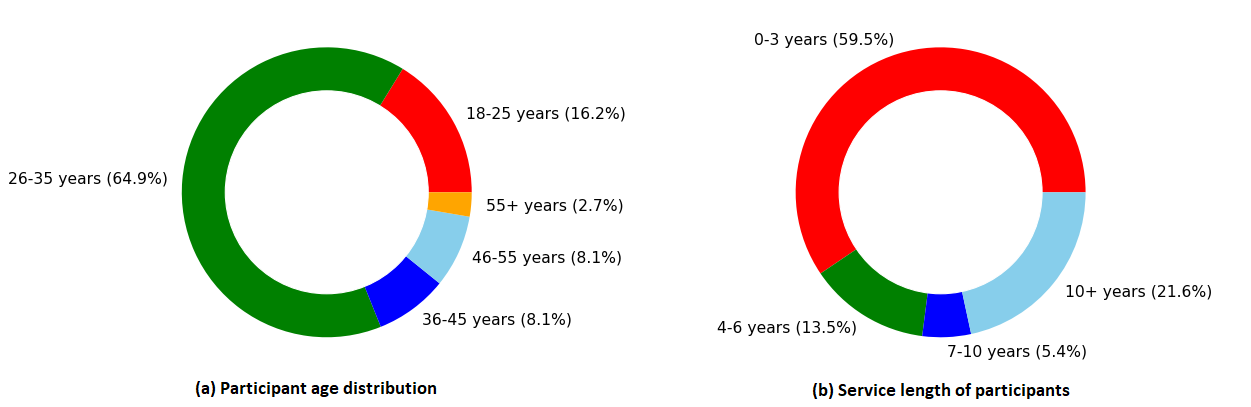}
  \caption{Participant statistics}
   \label{fig:participantStat}
\end{figure*}

The majority of previous research primarily investigated various pros and cons of workplaces, but the factors linked to potential limitations of \textit{hybrid work} settings caused by the real-world work conditions remains under-explored. For instance, due to the current COVID-19 crisis, workers may need to move between home offices and their traditional places of work maintaining staggered timing and social distancing rule imposed by many countries. This new normal forces workers to adjust to different workplace designs, layouts and facilities available which may incur a trade-off with workers' preferences such as preferred seating arrangement which is very crucial for many workers. In our recent research, we characterised concentration during work based on physical and ambient sensor data including air quality, temperature, humidity, movement, and noise ~\cite{9097829}. We also found that the unavailability of preferred seating arrangement at workplace can increase perceived stress and reduce concentration levels of workers while at work. But what preferred seats look like to different people was not characterised. Also, the meaning of preferred seating arrangement is not heavily investigated by recent literature. Therefore, the identification and implications of seating preferences when people choose to work in a hybrid work setting due to COVID-19 crisis is very timely to analyse.  

In this study, we investigate what a \textit{`preferred seating'} might look like and discuss a set of constraints associated with work productivity, by analyzing a dataset collected pre-COVID-19 from participants with diverse job roles. We present a list of implications leveraging our learning from the past. We also discuss how modern hybrid workplaces would be challenged and could overcome some of the key challenges while meeting these seating preferences and productivity constraints at work, especially given the current and post-COVID-19 crisis situations. 

The organization of this paper is as follows: Section 2 discusses the dataset and analyses seating preferences along with constraints of work productivity; Implications of our research  are discussed in Section 3; Finally, the paper concludes in Section 4.

\section{Seating and Productivity at Work}
In this section, we discuss a diverse set of seating preferences and several constraints that could influence workers' productivity. To analyse seating preference and productivity constraints, we built a dataset of 37 volunteer participants at Arup, Melbourne. The dataset was collected in May 2019. This research was approved by Human Research Ethics Committee of RMIT university, Australia. For data collection, we designed a survey in consultation with professional behavioural scientists. We logged responses by asking participants about their preferences related to seating arrangements while performing their work-related tasks. The participants also reported what influenced their perceived work productivity. The participant statistics and responses are analysed in the following subsections. Note that the pilot site is a traditional open-plan office who practices activity-based working. Since this is not a cubicle or room based office, there is no boundary among seats. Hence, many of the analysis could potentially be applicable to the home office setting as well. 

\subsection{Participant statistics}
The participants in our collected dataset are from a diverse professional backgrounds including mechanical engineers, structural engineers, building scientists, electrical engineers, acoustics and lighting professionals. In our participant cohort, the male and female proportions are 56.75\% and 43.24\%, respectively. As shown in Figure~\ref{fig:participantStat}(a), they are aged between 18 and 65 years, with a major portion of participants from 26-35 years age group. The length of service as a professional was also captured. As illustrated in Figure~\ref{fig:participantStat}(b), the service length of a large proportion of participants (i.e. 59.46\%) is under three years. A proportion of 21.6\% participants reported that they have worked for more than 10 years in this organization.

\subsection{Analyzing seating preferences}

A list of seating preferences is given to the participant to select, and they are also allowed to specify their preferences if it was not in the list. 
\begin{figure}[h]
  \centering
  \includegraphics[width=\columnwidth]{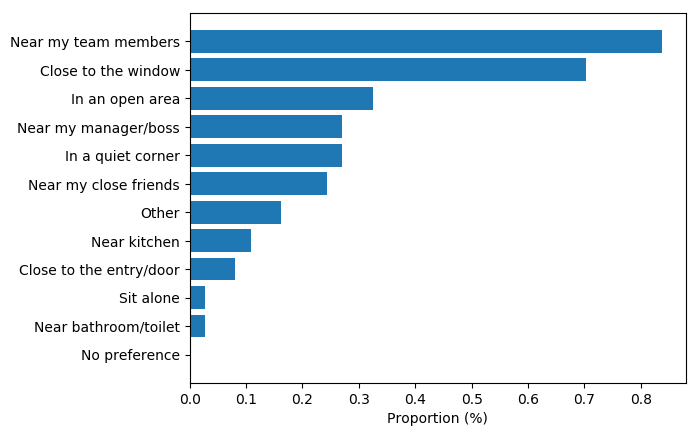}
  \caption{Distribution of seating preferences}
   \label{fig:Seating_preference}
\end{figure}

As shown in Figure~\ref{fig:Seating_preference},
%
all participants indicated that they have at least one preference, as ``\emph{no preference}" contributes 0\%. We found that 83.78\% of participants wanted to sit close to their team members. This is likely due to the fact that these workers need to interact with their team members while performing their tasks on specific projects. Many organisations allocate specific zones for the workers from same team. However, organizations and governments in many countries have imposed several restrictions to stop the spread of COVID-19 and to ensure safe-work for all employees. For instance, staggered start and finish times for businesses including schools have been in place in Australia to avoid crowding situations on public transport and workplaces \cite{vicgov1}. If employees need to do hot-desking or share common spaces, frequent cleaning needs to be carried out along with allowing only one person in every 4$m^2$ space\footnote{https://www.safeworkaustralia.gov.au/collection/workplace-checklists-covid-19}. Given the decreased room capacity due to social distancing and staggered timing requirement caused by COVID-19 crisis, having the all team member being present in the room or nearby would be much more challenging. To support staggered times of entry and exit to/from the workplaces, future research could develop solutions by formulating this problem as a new combinatorial optimisation problem. Another direction of research is to develop personalised recommender systems for suggesting a group of workers requiring entry to the workplaces based on their specific work requirements.

Next, ``\emph{close to the window}" and ``\emph{in an open area}" followed in preference, which accounted for more than 70.27\% and 32.43\% of responses, respectively. This is likely driven by the preference that many workers prefer to have some amount of natural lights during their working hours \cite{GALASIU2006728}. Many modern offices are designed to provide its occupants with enough open areas. However, offices may find it  challenging to arrange seating near the windows if the number of people with such preference is very large. In our dataset, a significant proportion of workers preferred to sit close to their managers or close friends. This may be important for some workers, however, it is not always easy to allocate workers with such seating arrangements due to the similar reasons as stated above. If someone needs to work from home, this closeness with team members may not be possible physically, however, could be replaced with technologies such as online chatroom.

A small proportion of participants (16.22\%) chose to specify their preferences, which is called as ``\emph{Other}". These includes adjustable sit/stand desk, desks with one or more large screens, and desks with good/clean keyboard and mouse. These may indicate the variable nature of preferences that could be very important to conduct some specific tasks as part of the nature of their job, or a requirement from an occupational health and safety point of view. For instance, one engineer specified seats ``\emph{wider/multiple screens}", which might be because they need to perform extensive drafting and analysis and feel comfortable with wider screens or a seating that provides multiple screens. Participants might also have pre-existing medical conditions that mean that height adjustable desks are required to ensure they don't get injured, or worsen their existing conditions. 

\subsection{Analyzing productivity constraints}

\begin{figure}[h]
  \centering
  \includegraphics[width=\columnwidth]{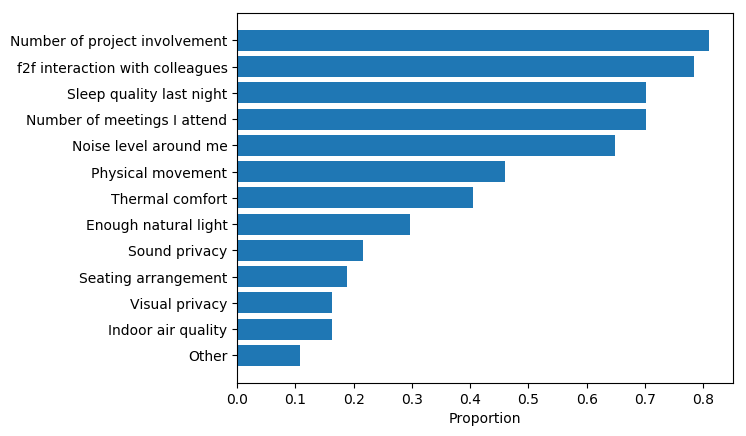}
  \caption{Distribution of productivity constraints}
   \label{fig:Productivity_constraints}
\end{figure}

Figure~\ref{fig:Productivity_constraints} shows the participants' identification of constraints to their work productivity. It is found that seating arrangements can influence the work productivity of workers. 
Specifically, 18.92\% of our participant workers reported that preferred seating arrangements effect their productivity at work. Another important point which could directly be linked to seating arrangement is the face to face communication with colleagues during work hours. In our participant cohort, 78.38\% reported that there is a direct association between face to face communication with colleagues and their work productivity. 

In hybrid workplaces, noise can have a negative impact towards work productivity. Open-plan offices are usually noisier than traditional office arrangements as reported by previous literature \cite{JAHNCKE2011373}.  Surrounding noise can cause distractions, and have an impact on the overall perceived concentration abilities of people \cite{rugg2010does,doi:10.1080/00140130412331311390}. In a ``\emph{working from home}" context, it is reasonable to assume that distractions with kids playing around, or family members doing other activities would have the same impacts on concentration. Figure \ref{fig:Productivity_constraints} shows that 64.86\% participants in our study reported this issue as a constraint to their work-related productivity.

\section{Current and Future Implications}

Seating arrangements and open plan offices have a direct impact on workers' productivity at work. Given the cost of decreased productivity and benefits of improved productivity, it is important to investigate how future workplaces can better manage constraints and enable drivers of concentration and productivity. 

From a behavioural point of view, the collected dataset highlights the overwhelming importance of proximity to team members and remaining connected to teammates as well as being located close to windows. This suggests that the success of workplace design relies on improving natural light intake and perimeter zone seating, as well as an operational management strategy to ensure desks are allocated appropriately for teams within an organisation. In other words, an open plan office arrangement whereby anyone can sit anywhere as they please, has potentially negative impacts on productivity by not controlling team arrangements. 

Multiple distributed factors that do not represent a majority indicate the importance of designing workplaces that have varied environments to meet individual demands. For example, although only 8\% of participants highlighted the importance of sitting close to the entry/door, it is important to cater for this minority, to ensure an inclusive workplace that aims to maximise ideal conditions for all workers and not disadvantage minority preference groups. Similarly, work spaces should be designed to cater for the 2.7\% of participants that prefer to sit alone. 

Future workplaces must also consider and design for limiting productivity constraints. The dataset shows that in general, environmental factors - apart from noise - are not as significant as operational and work-life balance factors. That is, productivity constraints are more associated to how meetings and project resourcing are managed, rather than comfort factors. In addition to these design factors, the dataset suggests the importance of human resource management and developing structures and systems to limit project numbers and meetings for each worker. 

In a post-COVID-19 society, it is likely that working from home will become more commonplace. Ensuring home-based seating configurations are optimised for each employee will become critical for managing productivity despite employers having limited control over these arrangements. This research has identified factors related to seating preferences and productivity at work using a dataset collected in pre-COVID-19 times that can be actively considered to address this issue. Regular team meetings via video conference for connectivity and human resource/project management have been identified as critical enablers and barriers to productivity. Ultimately, working from home arrangements can create the flexibility and opportunity for employees to tailor their spaces beyond normal levels. Therefore, hybrid workplaces where more people will continue working from in the future can presents an opportunity for employers to improve their workers’ concentration and productivity levels beyond what is possible in a commercial office. Individual workers can track their own constraints using latest technologies such as pervasive sensing based data collection and machine learning of their physical work environments \cite{9097829}. 

Moreover, the research presented in this paper will direct to future research in the areas of workplace recommender systems, constraint-based optimisation of work related preferences, and human-computer interactions which would lead the development of new interfaces for digital and physical workplaces.

\section{Conclusion}

This paper discusses various seating preferences and highlighted how it can impact the work related productivity of workers in a pre-COVID-19 workplace. From the analysis of our dataset collected from 37 participant workers, we found that everyone had at least one seating preference. A majority of the participants reported that a lack of seating arrangement of their preference can directly affect their work experience and overall productivity. We also discussed a list of current and future implications of our findings including the need for adaptation to home-based seating arrangements through seating configuration optimisation as working from home will be more common during and post-COVID-19 society.

We also discussed directions to future research by developing combinatorial optimisation solutions for staggered entry and exit to/from workplaces. This solution could be leveraged to develop efficient group recommendations for staggered entry to the workplaces. Future research also could investigate the effectiveness of adaptive seating preferences through the identification of employees' concentration, stress and productivity to provide better work support. To better understand individual requirements and enhanced work productivity, pervasive sensing and machine learning technologies developed for traditional workplaces could be transferred to work from home settings.

\begin{acks}
This research was supported by Arup and RMIT Enabling Capability Platform through the provision of an 'Opportunity Funding' scheme (no. 17073).
\end{acks}

\balance

\bibliographystyle{ACM-Reference-Format}
\bibliography{sample-base}

\end{document}